\documentclass[%
 reprint,
superscriptaddress,
 amsmath,amssymb,
aps,
prb,
]{revtex4-1}
\usepackage{graphicx}  
\usepackage{dcolumn}   
\usepackage{bm}        
\usepackage{verbatim}  
\usepackage{sidecap}

\usepackage{physics}
\usepackage{float}
\usepackage{color}
\usepackage[dvipsnames]{xcolor}
\usepackage[colorlinks]{hyperref}

\makeatletter
\@addtoreset{equation}{myequation}
\@addtoreset{figure}{myfigure}
\makeatother

\def \dag{^{\dagger}}
\def \ham{\mathcal{H}}

\newcommand{\vect}[1]{\mathbf{\bm{#1}}}

\newcommand{\vsigma}{\vect{\sigma}}

\def \c{\mathbf{c}}

\setcitestyle{numbers,square}

\begin{document}

\title{Topological {states} on the breathing kagome lattice}
\author{Adrien Bolens}
\affiliation{Institute of Physics, \'Ecole Polytechnique F\'ed\'erale de Lausanne (EPFL), 1015 Lausanne, Switzerland}
\affiliation{Department of Physics, University of Tokyo, Bunkyo-ku, Tokyo 113-0033, Japan}
\author{Naoto Nagaosa}
\affiliation{RIKEN Center for Emergent Matter Sciences (CEMS), Wako, Saitama 351-0198, Japan}
\affiliation{Department of Applied Physics, University of Tokyo, Bunkyo-ku, Tokyo 113-8656, Japan}

\date{\today}
\begin{abstract}
We theoretically study the topological properties of the tight-binding model on the breathing kagome lattice with antisymmetric spin-orbit coupling (SOC) between nearest neighbors. 
We show that the system hosts nontrivial topological phases even without second-nearest-neighbor hopping, and that the weakly dispersing band of the kagome lattice can become topological.
The main results are presented in the form of phase diagrams, where the $\mathbb{Z}_2$ topological index is shown as a function of SOC (intrinsically allowed and Rashba) and lattice trimerization. In addition, exact diagonalization is compared with effective low-energy theories around the high-symmetry points. We find that the weakly dispersing band has a very robust topological property associated with it. Moreover, the Rashba SOC can produce a topological phase rather than hinder it, in contrast to the honeycomb lattice.
Finally, we consider the case of a fully spin polarized 
(ferromagnetic) system, breaking time-reversal symmetry. We find a phase diagram that includes systems with finite Chern numbers. In this case too, the weakly dispersing band is topologically robust to trimerization.
\end{abstract}

\maketitle
\section{Introduction}
A series of theoretical predictions and experimental observations of topological band insulators (TBIs) has stimulated the emergence of a wide collection of topological quantum materials and topological phenomena \cite{hasan2010colloquium,bernevig2013topological}. The topological phases are classified according to an invariant in the bulk, as was originally shown for the noninteracting integer quantum Hall effect \cite{thouless1982quantized}, whereas gapless excitations appears at the border between topologically distinct phases.
Time-reversal invariant nonmagnetic insulators have been shown to reveal a topological $\mathbb{Z}_2$ classification which divides them into two categories described by the $\mathbb{Z}_2$ topological invariant $\nu$: trivial insulators ($\nu =0$), adiabatically connected to a trivial state, and ``topological" insulators ($\nu =1$) that are not connected to a trivial state without a bulk gap closure \cite{kane2005z2topological,kane2005quantum}.
In the original Kane-Mele model, intrinsically allowed spin-orbit coupling (SOC) between second nearest neighbor was shown to be essential to achieve a topological nontrivial phase \cite{kane2005z2topological}. In addition, both the inclusion of Rashba SOC, which breaks the reflection symmetry across the plane of the two-dimensional (2D) system, and of an inversion asymmetric on-site potential drive the insulator in the trivial phase when sufficiently strong.

The topological properties of the tight-binding model on the kagome lattice have also been investigated in the past \cite{liu2009spin,guo2009topological,zhigang2010quantum,liu2010simulating,zhang2011thequantum,tang2011high-temperature,wang2013prediction}. In particular, 2D organometallic topological insulators have been predicted for both the hexagonal and kagome lattices \cite{wang2013prediction, liu2013flat, wang2013organic, wang2013quantum}. Recently, coordination nanosheets with atomic thickness have been intensively studied both experimentally and theoretically because of their attractive physical and chemical properties \cite{maeda2016coordination}. The synthesis of such materials is an exciting way of potentially engineering new materials with nontrivial topological properties. 
{In addition, on the kagome lattice, topological properties of Floquet-Bloch band structures have been studied in Floquet systems \cite{du2017Quadratic}, and the magnon bands of kagome magnets with SOC have been shown to be topological and lead to a finite magnon thermal Hall conductivity \cite{zhang2013topological, chisnell2015topological, laurell2018magnon}.
}

 On the kagome lattice, in contrast to the honeycomb one, there is no inversion symmetry centered on the middle of the nearest-neighbor bonds, and antisymmetric SOC is allowed between nearest neighbors \cite{dzyaloshinsky1958thermodynamic,moriya1960anisotropic}. Kagome systems with such intrinsic SOC between nearest neighbors were shown to host nontrivial phases \cite{liu2010simulating,tang2011high-temperature, wang2013prediction}. Moreover, the energy spectrum of the tight-binding model on the kagome lattice has an extra dispersion-less band which can also become topological thanks to SOC 
\cite{liu2010simulating}.
Topological flat bands have been studied in the context of the fractional quantum Hall effect and topological flat-band lattice models \cite{neupert2011fractional,sun2011nearly,tang2011high-temperature,yang2012topological}, and are an important feature of some TBIs.

In real materials, the mirror symmetry about the plane containing the 2D system is often not protected, resulting in Rashba SOC between nearest neighbors, in addition to the intrinsic SOC. For instance, the breaking of the mirror symmetry can be caused by the interaction with a substrate, by buckling of the 2D material, or simply by an external electric field \cite{liu2011low-energy, wang2013organic}. In addition, inversion symmetry-breaking lattice trimerization (i.e., the breathing kagome lattice) can also be important for materials on the kagome lattice \cite{aidoudi2011anionothermally, orain2017nature}.

In this paper, we thus study systems with both types of SOC and with the lattice trimerization. We show that the inclusion of Rashba SOC drives the system into a topological phase by itself, in contrast to graphene where it hinders the nontrivial phase. In addition, we show that the flat band is intrinsically topological, in a very robust way.
We note that the trimerization term has already been considered in Ref.~\cite{liu2010simulating}, but only perturbatively, and not in combination with Rashba SOC.
%

The present paper is structured as follows. First, in Sec.~\ref{sec:model}, we introduce the model considered.
We then calculate the $\mathbb{Z}_2$ topological index in Sec.~\ref{sec:Z2} and draw the full phase diagram considering nearest-neighbor SOC (intrinsic and Rashba) and lattice trimerization. 

Subsequently, in Sec.~\ref{sec:chern}, we consider the same model but for a system with ferromagnetically polarized spins and derive an effective three-band tight-binding model. We calculate the Chern numbers of the three bands and draw the phase diagram accordingly.

Finally, we conclude in Sec.~\ref{sec:conclusion}.

\begin{figure}
\centering
	\includegraphics[width=0.35\textwidth]{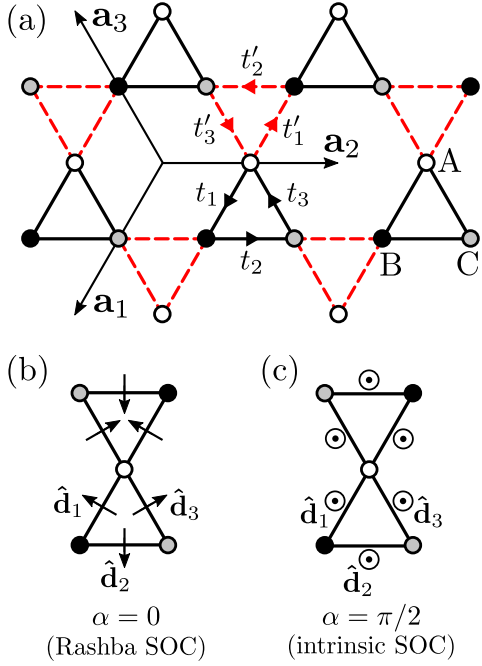}
	\caption{(a) Breathing kagome lattice with the hopping terms of Eqs.~\eqref{eq:ti} and \eqref{eq:t'i}
	indicated and the definition of the lattice vectors $\vb a_1$, $\vb a_2$, and $\vb a_3$. White, black, and gray dots represents sites in the A, B, and C sublattices, respectively. (b) SOC vectors $\vu d_1$, $\vu d_2$, and $\vu d_3$ defined in Eq.~\eqref{eq:di} for pure Rashba SOC ($\alpha =0$). (c) Same SOC vectors for pure intrinsic SOC ($\alpha = \pi/2$).}
\label{fig:kagome}
\end{figure}

\section{Model}
\label{sec:model}
Let us consider the tight-binding model on the kagome lattice as depicted in Fig.~\ref{fig:kagome}. The spin-independent Hamiltonian is written as
\begin{align}
\label{eq:ham1}
  \ham_1 = - \sum_{\vb R}[
    &t\c\dag_{\vb R B} \c_{\vb R A} + t'\c\dag_{(\vb R - \vb a_1) B} \c_{\vb R A} \nonumber \\
  + &t\c\dag_{\vb R C} \c_{\vb R B} + t'\c\dag_{(\vb R - \vb a_2) C} \c_{\vb R B} \nonumber \\
  + &t\c\dag_{\vb R A} \c_{\vb R C} + t'\c\dag_{(\vb R - \vb a_3) A} \c_{\vb R C}] + \rm{H.c.}, 
\end{align}
where $\mathbf{c}_{\vb R \alpha} \dag=(c\dag_{\vb R \alpha \uparrow}, c\dag_{\vb R \alpha \downarrow})$ with $\alpha \in \{A, B, C\}$ for the three sublattices. Here $\vb R$ labels the position of the unit cell composed of three sites forming an upward triangle, and $\vb a_1 = (-1/2,-\sqrt{3}/2)$, $\vb a_2 = (1,0)$, and $\vb a_3 = (-1/2, \sqrt{3}/2)$ [see Fig.~\ref{fig:kagome}(a)]. The trimerization of the lattice, corresponding to the breathing kagome lattice, is obtained by considering different hopping amplitudes on the two distinct sets of triangles: $t = t_0 + \delta$ for upward triangles and $t' = t_0 - \delta$ for downward triangles. Here $\delta$ is trimerization parameter, and \eqref{eq:ham1} can be decomposed as $\ham_1 = \ham_0(t_0) + \ham_{\rm trim}(\delta)$.

The SOC coupling is introduced as a spin-dependent hopping between nearest neighbors. Furthermore, we assume that SOC hopping amplitudes are similarly affected by the trimerization of the lattice. The general nearest-neighbor SOC Hamiltonian is
\begin{align}
  \ham_{\rm SO} =i\sum_{\vb R}[ &\lambda \c\dag_{\vb R B} (\vu d_1 \cdot \vsigma) \c_{\vb R A} + \lambda'\c\dag_{(\vb R - \vb a_1) B} (\vu d_1 \cdot \vsigma) \c_{\vb R A} \nonumber \\
  + &\lambda \c\dag_{\vb R C} (\vu d_2 \cdot \vsigma) \c_{\vb R B} + \lambda'\c\dag_{(\vb R - \vb a_2) C}(\vu d_2 \cdot \vsigma) \c_{\vb R B} \nonumber \\
  + &\lambda\c\dag_{\vb R A} (\vu d_3 \cdot \vsigma)\c_{\vb R C} + \lambda'\c\dag_{(\vb R - \vb a_3) A} (\vu d_3 \cdot \vsigma)\c_{\vb R C}] \nonumber \\
  +& \rm{H.c.}, 
\end{align}
with $\lambda = \lambda_0 + \delta_{\lambda}$ and $\lambda' = \lambda_0 - \delta_{\lambda}$ so that $\lambda/\lambda' = t/t'$ (i.e., $\delta_{\lambda} = \lambda \cdot \delta/t_0$). Here, $\vsigma = (\sigma_x, \sigma_y, \sigma_z)$ are the Pauli matrices and the $\vu d_i$'s are unit vectors which depend on the type of SOC considered. The intrinsic SOC (i.e., it respects all symmetries of the lattices including the mirror symmetry across the 2D plane) corresponds to $\vu d_1 = \vu d_2 = \vu d_3 = \vu z$ [see Fig.~\ref{fig:kagome}(c)] with amplitudes denoted by $\lambda = \lambda_{\rm i} = \lambda_{0, \rm i} + \delta_{\lambda, \rm i}$ and $\lambda' = \lambda'_{\rm i} = \lambda_{0, \rm i} - \delta_{\lambda, \rm i}$. In a system breaking the mirror symmetry, we also have Rashba SOC with amplitudes denoted by $\lambda = \lambda_{\rm R} = \lambda_{0, \rm R} + \delta_{\lambda, \rm R}$ and $\lambda' = \lambda'_{\rm R} = \lambda_{0, \rm R} - \delta_{\lambda, \rm R}$, which corresponds to in-plane $\vu d_i$'s perpendicular to their respective bonds, as depicted in Fig.~\ref{fig:kagome}(b). We consider the general situation with both types of SOC, so that
\begin{align}
  \vu d_1 &= \cos \alpha (\frac{\sqrt{3}}{2}, -\frac{1}{2},0) + \sin \alpha (0,0,1), \nonumber \\
  \vu d_2 &= \cos \alpha (0,1,0) + \sin \alpha (0,0,1),\nonumber \\
  \vu d_3 &= \cos \alpha (-\frac{\sqrt{3}}{2}, -\frac{1}{2},0) + \sin \alpha (0,0,1),
  \label{eq:di}.
\end{align}
Here $\alpha$ is the angle between the 2D plane and the $\vu d_i$'s such that $\alpha=0$ corresponds to pure Rashba SOC and $\alpha=\pi/2$ to pure intrinsic SOC: $\tan \alpha = \lambda_{0,\rm i}/\lambda_{0,\rm R}$.

The full Hamiltonian $\ham = \ham_0 + \ham_{\rm trim} + \ham_{\rm SO, i} + \ham_{\rm SO, R}$ is
\begin{align}
  \ham =-\sum_{\vb R}[ & \c\dag_{\vb R B} \hat t_1 \c_{\vb R A} + \c\dag_{(\vb R - \vb a_1) B} \hat t'_1 \c_{\vb R A} \nonumber \\
  + & \c\dag_{\vb R C} \hat t_2 \c_{\vb R B} + \c\dag_{(\vb R - \vb a_2) C}\hat t'_2 \c_{\vb R B} \nonumber \\
  + & \c\dag_{\vb R A} \hat t_3 \c_{\vb R C} + \c\dag_{(\vb R - \vb a_3) A} \hat t'_3\c_{\vb R C}] + \rm{H.c.}, 
\label{eq:full}
\end{align}
where each hopping is accompanied by a spin rotation defined by
\begin{align}
\label{eq:ti}
  \hat t_i &= \sqrt{t^2 + \lambda_{\rm i}^2 + \lambda_{\rm R}^2} \, \exp(i\phi \vu d_i \cdot \vsigma),  \\
\label{eq:t'i}
  \hat t'_i &= \sqrt{t'^2 + \lambda_{\rm i}'^2 + \lambda_{\rm R}'^2}\,  \exp(i\phi \vu d_i \cdot \vsigma), \\
  \phi &= -\arctan(\frac{\sqrt{\lambda_{0, \rm i}^2 + \lambda_{0,\rm R}^2}} {t_0}).
\label{eq:phi}
\end{align}

\begin{figure}
\centering
	\includegraphics[width=0.5\textwidth]{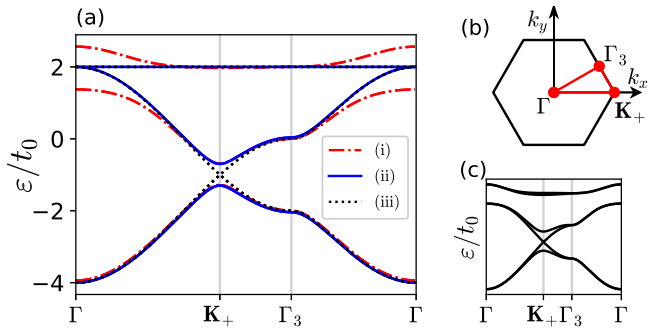}
	\caption{(a) Band structure with pure Rashba SOC ($\alpha = 0$) with (i) $\delta = 0, \lambda_{\rm 0, R} = 0.055 t_0$, (ii) $\delta = 0.1 t_0, \lambda_{0,\rm R} = 0$, and (iii) $\delta = \lambda_{\rm 0, R} = 0$. (b) Brillouin zone with high-symmetry points. (c) Band structure with $\delta = 0.1 t_0$ and $\lambda_{\rm 0, R} = 0.055 t_0$ which corresponds to a band-touching at the $\vb K_{\pm}$ points.}
\label{fig:disp}
\end{figure}

After a Fourier transform $\c\dag_{\vb R \alpha} = \frac{1}{\sqrt{N}}\sum_{\vb k}\c\dag_{\vb k \alpha} e^{-i \vb k \cdot \vb R}$, where $N$ is the number of unit cells, we obtain the Bloch Hamiltonian $\ham(k)$. In the $(\c\dag_{\vb k A},\c\dag_{\vb k B},\c\dag_{\vb k C})$ basis, the $6 \times 6$ matrix is
\begin{equation}
\label{eq:bloch}
  \ham(k) = -\begin{pmatrix}
  	0 & \hat t_1\dag + \hat t_1'^{\dagger} e^{-i \vb k \cdot \vb a_1} &\hat t_3 +  \hat t_3' e^{i \vb k \cdot \vb a_3} \\
  	\hat t_1 + \hat t_1' e^{i \vb k \cdot \vb a_1} & 0 & \hat t_2\dag + \hat t_2'^{\dagger} e^{-i \vb k \cdot \vb a_2} \\
  	\hat t_3\dag + \hat t_3'^{\dagger} e^{-i \vb k \cdot \vb a_3} & \hat t_2 + \hat t_2' e^{i \vb k \cdot \vb a_2} & 0
  \end{pmatrix}.
\end{equation}
Without SOC nor trimerization, the Hamiltonian~\eqref{eq:bloch} has three doubly degenerate bands: two pairs of dispersing bands similar to graphene with two Dirac points at $\vb K_{\pm}=(\pm 4\pi/3,0)$ and an additional pair of flat bands as shown in Fig.~\ref{fig:disp}(a)(iii).

A gap opens at the Dirac points when introducing either SOC or trimerization.
 However at the quadratic band touching point, a gap does not open from trimerization alone but only opens when either type of SOC is introduced.
In this case, the top bands are not perfectly flat anymore, but are only weakly dispersing as long as SOC is small ($\lambda_0 \ll t_0$), as can be seen in Fig.~\ref{fig:disp}(a)(i) and \ref{fig:disp}(c).

 We also note that the transformation $\phi \rightarrow \phi \pm \pi$ is equivalent to $t_i \rightarrow -t_i$ and $t'_i \rightarrow -t'_i$ so that the topological properties at 1/3 filling with angle $\phi$ are the same as at 2/3 filling with angle $\phi \pm \pi$.

\section{$\mathbb{Z}_2$ topological invariant}
\label{sec:Z2}
\begin{figure*}
\centering
	\includegraphics[width=0.85\textwidth]{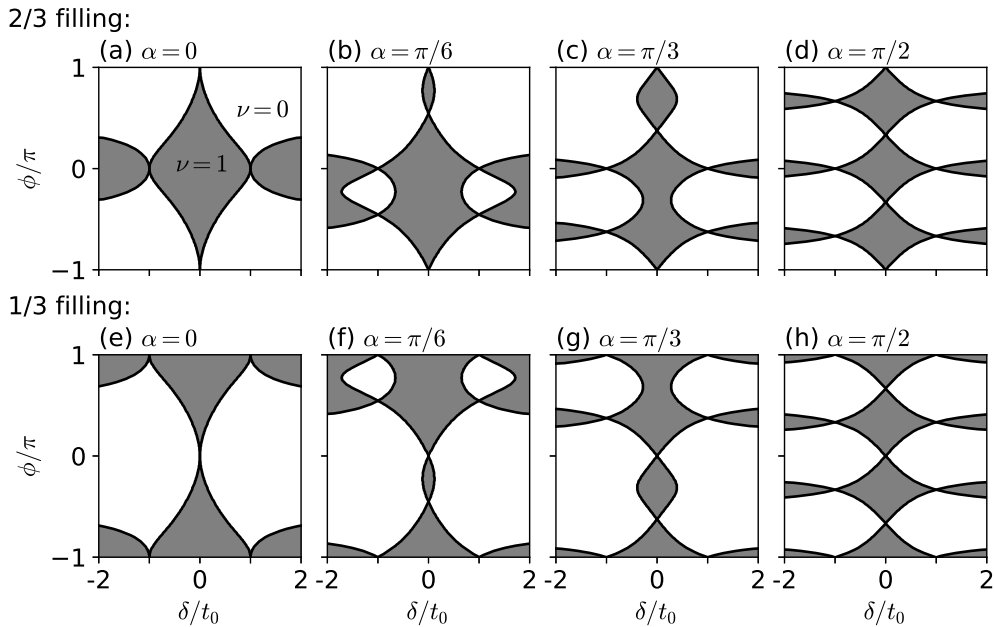}
	\caption{Phase diagrams as a function of $\delta/t_0$ and $\phi$ for various SOC vector directions corresponding to $\alpha = 0$ (pure Rashba), $\pi/6$, $\pi/3$, and $\pi/2$ (pure intrinsic). The $\nu = 0$ and $1$ regions are indicated in white and gray, respectively. Both 2/3 filling (top) and 1/3 filling (bottom) cases are shown, which are related by a $\phi \rightarrow \phi \pm \pi$ transformation.}
\label{fig:phase_dia}
\end{figure*}

\subsection{Results}
First we briefly discuss the topology of the system {without trimerization, $\delta = 0$}. In this case, the system is invariant under inversion symmetry and the $\mathbb{Z}_2$ topological invariant $\nu$ is easily obtained from the parity eigenvalues $\xi_{2m}(\Gamma_{1-4})$ of the $2m$th occupied bands with $m=1,2$, or $3$ at the four time-reversal-invariant ($\mathcal{T}$-invariant) points $\Gamma_{1}=\Gamma, \Gamma_2=(2\pi,0), \Gamma_3 = (\pi, \pi/\sqrt{3}), \Gamma_4 =(\pi, -\pi/\sqrt{3})$ \cite{fu2007topological}. For the kagome lattice, the parity operator at the $\mathcal{T}$-invariant points, which is spin-independent, is concisely written as a $3 \times 3$ matrix,
\begin{equation}
  \mathcal{P}(\vb k) = {\rm diag}(1, e^{i \vb a_1 \cdot \vb k},  e^{-i \vb a_3 \cdot \vb k}),
\end{equation}
where we chose the inversion center on an A site so that $\mathcal{P}(\Gamma_1) = (1,1,1)$, $\mathcal{P}(\Gamma_2) = (1,-1,-1)$, $\mathcal{P}(\Gamma_3) = (1,-1,1)$, and $\mathcal{P}(\Gamma_4) = (1,1,-1)$. The $\mathbb{Z}_2$ index for the $m$th Kramers pair is given by $(-1)^{\nu_m}= \prod_{i} \xi_{2m}(\Gamma_i)$ and, by explicitly calculating the wave functions of the Bloch Hamiltonian~\eqref{eq:bloch}, we find that $\nu_m=1$ for both the bottom and top pairs of bands independently of $\phi$. Hence, the overall $\mathbb{Z}_2$ index is always $\nu = 1$ for both 1/3 and 2/3 filling (as long as the different Kramers pairs of bands are not touching). Most interestingly, this is not only true for the intrinsic SOC, but also for the Rashba SOC.

Motivated by this preliminary result, we show the full phase diagram in Fig.~\ref{fig:phase_dia} as a function of both $\phi$ and the inversion symmetry-breaking $\delta$ for several values of $\alpha$. The phase diagrams are plotted for $\alpha =0$ (pure Rashba), $\pi/6$, $\pi/3$, and $\pi/2$ (pure intrinsic) considering both 1/3 and 2/3 fillings. They were obtained by numerically calculating the number of zeros of the Pfaffian
\begin{equation}
  p(\vb k) = {\rm Pf}[\mel{u_i(\vb k)}{\mathcal{T}}{u_j(\vb k)}],
\end{equation}
which is the method originally described by Kane and Mele \cite{kane2005z2topological}. Here $\mathcal{T}$ is the time-reversal operator, $\ket{u_i(\vb k)}$ are the band wave functions, and $i$ ranges over the filled bands. Due to the $D_3$ symmetry of the breathing kagome lattice, only zeros along the high-symmetry $[\Gamma-\Gamma_2]$ line segment (along which $k_y=0$) are relevant. We observed that the band-touchings happen at either $\vb k = \vb K_\pm$ or $\vb k = \Gamma$. All gap-closing-gap-opening transitions at the $\vb K_{\pm}$ Dirac points (which always happen simultaneously for both points) result in a change of the topological index. When $\phi = 0$ or $\pi$, the pair of flat bands touches another dispersive pair of bands ``quadratically" at $\vb k = \Gamma$, but the topological index is not affected. Finally, in the unphysical case where $t = -t'$ (or $\delta \rightarrow \pm\infty$), all six bands are degenerate at the $\Gamma$ point and $\nu$ changes in a nontrivial way \footnote{This is not observed in Fig.~\ref{fig:phase_dia} because $t$ and $t'$ change sign when $\delta$ goes from $\infty$ to $-\infty$.}.

\subsection{Discussion}

In the following, we derive effective Hamiltonians in two different limits to give some analytical insight to the phase diagram: (1) the $\lambda_0, \delta \ll t_0$ limit (small SOC and trimerization) relevant at $1/3$ filling and (2) the $t' \ll t$ limit (close to full trimerization $\delta/t_0 =1$), relevant at $2/3$ filling.

\subsubsection{Small SOC and trimerization}

The gap-closing-gap-opening behavior near the Dirac points, relevant at $1/3$ filling, can be understood by considering the linearized Dirac Hamiltonian expanded around $\vb k = \vb K_{\pm}$. The effective Hamiltonian is obtained after a projection onto the subspace of the two pairs of bands forming the Dirac cones. For the trimerization and intrinsic SOC,
\begin{align}
\label{eq:dirac}
  \ham_{\tau s \sigma}(\vb k) =&\, v_F \qty( k_x  \sigma_x \tau_z + k_y \sigma_y  )\nonumber \\
   &+ (\sqrt{3}\lambda_{0,{\rm i}} \tau_z s_z - 3 \delta )\sigma_z + \rm{const.},
\end{align}
with $v_F = \sqrt{3}t_0/2$ and $\tau_i$, $s_i$, and $\sigma_i$ ($i \in \{x,y,z \}$) are Pauli matrices which refer to the $\vb K_{\pm}$ valleys, spins, and bands (in a given basis), respectively. The constant term in Eq.~\eqref{eq:dirac} includes all terms independent of the $\sigma_i$'s. A similar result has already been obtained in Ref.~\cite{liu2010simulating}. We see from Eq.~\eqref{eq:dirac} that the band at $1/3$ filling becomes topological when $\abs{\lambda_{0, {\rm i}}} > \sqrt{3} \abs{\delta}$.

For the $\lambda_{0, \rm R}$ Rashba term, we obtain no contribution coming out of this projection, which apparently contradicts what we observe numerically in Fig.~\ref{fig:phase_dia}.
However, in order to obtain Eq.~\eqref{eq:dirac}, the projection was done on the eigenstates at the $\vb K_{\pm}$ points calculated without SOC and with $\delta =0$. Hence, it corresponds to a first order perturbation theory in $\ham_{\rm trim}$, $\ham_{\rm SO, i}$, and $\ham_{\rm SO, R}$. The first contribution from the Rashba SOC comes at second order (in $\lambda_{\rm 0,R}$ and $\delta$) and we find
\begin{equation}
  \ham_{\rm SO, R} \rightarrow  \sqrt{3} \delta_{\lambda,{\rm R}} (\sigma_x s_y \tau_z - \sigma_y s_x) + \frac{\lambda_{\rm 0, R}^2}{2t_0} \tau_z s_z \sigma_z,
  \label{eq:HamR}
\end{equation}
where we can also write $\delta_{\lambda,{\rm R}} =  \lambda_{\rm 0, R} \delta/t_0$.
Hence, Rashba SOC drives the system in topological phase by itself when $\lambda_{\rm 0, R}^2 > 6 |\delta \,  t_0| + \mathcal{O}(\delta^2)$. Both the linear and quadratic behaviors can be seen around the $(\delta/t_0 =0, \phi = 0$) point in Fig.~\ref{fig:phase_dia}(h) and Fig.~\ref{fig:phase_dia}(e), respectively.

The Hamiltonians~\eqref{eq:dirac} and \eqref{eq:HamR} have a limitation: They do not tell anything about the topology of the top bands. In particular, the gap opened at the {quadratic band touching point} $\Gamma$  is stable under the inclusion of {the trimerization} $\delta$, even when a gap closes at a Dirac point, as shown in Fig.~\ref{fig:disp}(c), which reflects the fact that $\delta$ by itself does not open a gap at $\Gamma$. 

%

\begin{figure}[t]
\centering
  \includegraphics[width=0.9\linewidth]{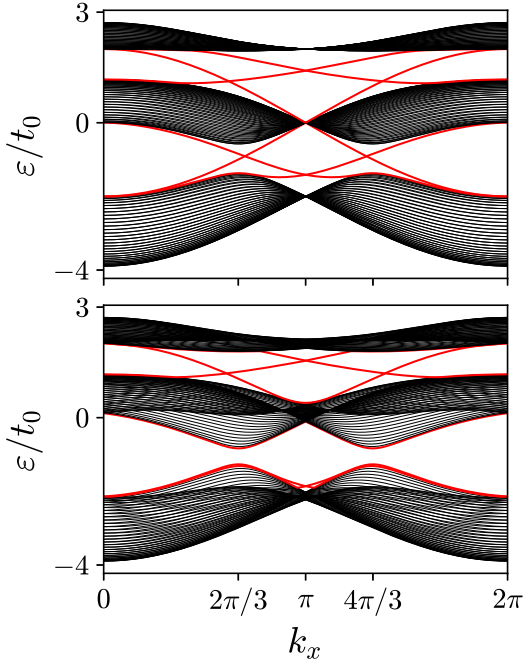}
\caption{Energy dispersion of a system periodic in the $\vu x$ direction and with edges in the $\vu y$ direction, with $\alpha = 0$ (Rashba SOC), $\phi = 0.3 \pi$, and (top): $\delta = 0.0$, (bottom): $\delta = 0.2$.}
\label{fig:edge}
\end{figure}

The full phase diagram reveals that the pair of {weakly dispersing} bands is intrinsically topological. As soon as the quadratic band touching at $\Gamma$ is lifted, e.g., by an infinitesimally small SOC contribution, the now well-defined $\mathbb{Z}_2$ index of the {top bands} is nontrivial until a band-closing-band-opening happens at $\vb K_{\pm}$ for large {trimerization or SOC: $\delta$ or $\lambda_0 \sim t_0$}. Thus, the topological phase at $2/3$ filling is very stable thanks to the  {weakly dispersing} bands and exists over a large range of values for $\delta$. As opposed to the $1/3$ filling case, an arbitrarily small SOC will gap the flat bands in a topological way independently of the strength of {the trimerization} $\delta$ (as long as $-1 < \delta/t_0 < 1$). In Fig.~\ref{fig:edge}, we show the energy dispersion of a system with edges in the $\vu y$ direction for $\alpha=0$ (Rashba SOC), $\phi=0.3\pi$, and both $\delta = 0$ (top) and $\delta =0.2 t_0$ (bottom). We clearly see the change in the edge states at $1/3$ filling from the topological phase at $\delta=0$ to the trivial phase at $\delta=0.2 t_0$, while at $2/3$ filling the phase stays topological.

\subsubsection{Fully trimerized limit}

We now derive a second effective Hamiltonian for large {trimerization} $\delta$, valid near the fully trimerized limit $\delta = t_0$, or $t'=0$. We also assume small SOC, $\lambda \ll t$.

The fully trimerized system consists of independent upward triangles. The now localized system has no energy dispersion, and the Hamiltonian is simply
\begin{equation}
  \ham(k) = -t \begin{pmatrix}
  	0 & 1 & 1 \\
  	1 & 0 & 1 \\
  	1 & 1 & 0
  \end{pmatrix} \otimes \mathbb{I}_{2},
  \label{eq:trimerH0}
\end{equation}
where $\mathbb{I}_{2}$ is the identity in spin space, so that there are two pairs of bands with energy $E= t$ and one pair with energy $E= -2t$. The effective theory at $2/3$ filling can thus be obtained by projecting the other Hamiltonians onto the subspace spanned by the $E=t$ states. The dispersion comes from the small hopping $t'$ between the nearly isolated triangles. Expanding $\ham_1$ {in Eq.~\eqref{eq:ham1}} with finite $t'$ around $K_\pm$ we obtain
\begin{equation}
  \ham_1 \rightarrow v_F' \qty( -k_x \sigma_z + k_y \sigma_x )\tau_z + \frac{3t'}{2} \sigma_y \tau_z,
  \label{eq:trimerH1}
\end{equation}
where $v_F' = \sqrt{3} t'/2$, and the $\sigma_i$ Pauli matrices now refer to the two $E=t$ bands [we have chosen two real eigenvectors of Eq.~\eqref{eq:trimerH0}].
The intrinsic SOC on each trimer gives a gap term,
\begin{equation}
  \ham_{\rm SO, i} \rightarrow  \sqrt{3} \lambda_{\rm i} \sigma_y s_z,
  \label{eq:trimerHi}
\end{equation}
and the contribution from Rashba SOC only comes at second order in $\lambda_{\rm R}$ and $t'$,
\begin{equation}
  \ham_{\rm SO, R} \rightarrow  \frac{\sqrt{3}}{2} \lambda_{\rm R}'\qty(\sigma_x s_x + \sigma_z s_y)\tau_z - \frac{\lambda_{\rm R}^2}{2t} \sigma_y s_z,
  \label{eq:trimerHR}
\end{equation}
where we can also write $\lambda_{\rm R}' = \lambda_{\rm R} t'/t$.

{When the intertriangle hopping $t'=0$, the completely localized system is obviously topologically trivial. This corresponds to the $\delta/t_0=1$ lines in Fig.~\ref{fig:phase_dia}. Then,
as $t'$ increases, the previously $E=t$ degenerate bands split in a topological manner.
For small SOC and $t'$ ($\lambda, t' \ll t$)}, the phase boundary is given by Eqs.~\eqref{eq:trimerH1}-\eqref{eq:trimerHR}: The system is topological if $ |\lambda_i| < \sqrt{3}|t'|/2 $ for pure intrinsic SOC, and $\lambda_R^2 < 3|t \, t'|$ for pure Rashba SOC. The linear and quadratic behavior can be seen around the $(\delta/t_0=1, \phi =0)$ point in Fig.~\ref{fig:phase_dia}(d) and Fig.~\ref{fig:phase_dia}(a), respectively.

It is interesting to note that, effectively, the trimerized system at $2/3$ filling can be seen as a two-band system on a triangular lattice. 
The nontrivial phase arises thanks to the complicated effective hopping Hamiltonian $\ham_1$ and the on-site two-band Hamiltonian obtained after projecting out the third pair of bands.

The effective model with, say, intrinsic SOC can be written on a triangular lattice (made of the original unit cells at positions $\vb R$) as
\begin{align}
  \ham_{\rm eff} =& \sum_{\vb R}\qty[ \sum_{i=1}^3  \vb f\dag_{\vb R + \vb a_i} (\hat T_i \otimes \mathbb{I}_2) \vb f_{\vb R} + \rm{H.c.} ]\nonumber \\
   &+ \sqrt{3} \lambda_{\rm i} \sum_{\vb R} \vb f\dag_{\vb R} (\sigma_y \otimes s_z) \vb f_{\vb R},
\end{align}
where $\vb f\dag_{\vb R} = (f_{\vb R, a, \uparrow}\dag, f_{\vb R, a, \downarrow}\dag, f_{\vb R, b, \uparrow}\dag, f_{\vb R, b, \downarrow}\dag)$,  $a, b= 1,2$ are band indices, and $f_{\vb R, a, s}\dag$ ($f_{\vb R, b, s}\dag$) is the appropriate linear combination of $c_{\vb R,\alpha, s}\dag$ ($\alpha = A,B,C$) corresponding to the $a$ ($b$) band.
The complexity of the model is hidden in the hopping matrices $\hat T_i$, which are different for the three directions $\vb a_1$, $\vb a_2$, and $\vb a_3$, and break the inversion symmetry on each bond (i.e., $\hat T_i\dag \neq \hat T_i$):
\begin{align}
  &\hat T_1 = \frac{t'}{3} \begin{pmatrix}
  	1 & -\sqrt{3} \\
  	0 & 0  \\
  \end{pmatrix}, \, \hat T_2 =  \frac{t'}{6} \begin{pmatrix}
  	-1 & -\sqrt{3} \\
  	\sqrt{3}  & 3  \\
  \end{pmatrix}, \nonumber \\
  &\hat T_3 = \frac{t'}{3} \begin{pmatrix}
  	1 & 0\\
  	\sqrt{3}  & 0  \\
  \end{pmatrix}.
\end{align}

\begin{figure}
\centering
	\includegraphics[width=0.4\textwidth]{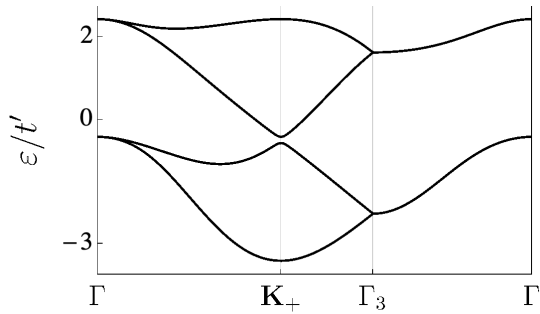}
	\caption{Energy dispersion of the effective Hamiltonian in the nearly-trimerized limit with SOC given in Eqs.~\eqref{eq:trimerH1} and \eqref{eq:trimerHi} near the phase transition: $\lambda_i \lesssim \sqrt{3}t'/2$.}
\label{fig:disptri}
\end{figure}

The band dispersion of this effective model is shown in Fig.~\ref{fig:disptri} near the band-closing-band-opening transition. Note that in the topological phase near the trimerized limit, there is no overall gap between the two pairs of band when including only linear terms in $t'/t$ (the third band in Fig.~\ref{fig:disptri} has exactly the same energy at $\vb k = \vb K_{\pm}$ as the lower bands at $\vb k=\Gamma$), but higher order terms in $t'/t$ will gap the two pairs of band completely (i.e., there is a full gap in the full model with all bands).

Finally, we note that in the unphysical $\delta/t_0 \notin [-1, 1] $ situation (i.e., the trimerization results in hopping $t$ and $t'$ with different signs), we can still define the topological index, but the system is in a semimetal phase with no overall gap, even in the full model.

\section{Spin-polarized system}
 \label{sec:chern}
 Let us finally consider a ferromagnetically ordered system where, on each site, the magnetic dipole points along the same unit vector $\vu e= (\sin \Theta \cos \Phi, \sin \Theta \sin \Phi, \cos \Theta)$. In the following, we show that the system can host nontrivial Chern numbers for the three bands, as shown in Ref.~\cite{ohgushi2000spin} for the $\delta=0$ case, and draw the resulting phase diagram for $\delta \neq 0$ in Fig.~\ref{fig:PD_chern}.
 
Such a constraint can be enforced in our system through the double-exchange model
\begin{equation}
 \ham_{\rm DE} = \ham - \mathcal{A} \sum_{\vb R} \sum_{\alpha = A,B,C} \vb c\dag_{\vb R \alpha}\qty( \vu e \cdot \vsigma) \vb c_{\vb R, \alpha},
\end{equation}
where $\ham$ is given in Eq.~\eqref{eq:full}. The coupling to the localized magnetic moments (with $\mathcal{A}>0$) typically originates from Hund's coupling, but it could also originates from an on-site repulsion $U$, based on the Hubbard model at the mean field level. For large $\mathcal{A}$, we effectively only keep the component of the spinor parallel to $\vu e$, $\ket{\chi_+}$, such that we are left with a three-band model with effective hopping
\begin{align}
  t_i^{\rm eff} &= \mel{\chi_+}{\hat{t}_i}{\chi_+} \equiv e^{i \varphi_i} |t_i^{\rm eff}|, \nonumber \\
  t_i^{\prime \textrm{eff}} &= \mel{\chi_+}{\hat{t}'_i}{\chi_+} \equiv e^{i \varphi_i} |t_i^{\prime \rm eff}|.
\end{align}
We then define $\varphi = \varphi_1 + \varphi_2 + \varphi_3$, and choose the gauge where $\varphi_1 = \varphi_2 = \varphi_3 = \varphi/3$.

Furthermore, for the sake of simplicity, we only consider cases where $|t_1^{\rm eff}|=|t_2^{\rm eff}|=|t_3^{\rm eff}|$, so that the $C_3$ symmetry is preserved. For intrinsic SOC, this is always the case and we have the relation $\tan(\varphi/3) = \cos(\Theta) \tan(\phi)$ {[where $\phi$ quantifies the SOC, as defined in Eq.~\eqref{eq:phi}]}, so that $\varphi$ is finite unless $\vu e$ lies in the plane. For Rashba SOC, we can only have the $C_3$ symmetry for $\Theta=0,\pi$, in which cases $t_i^{\rm eff} = t \cos(\phi)$, and $\varphi=0$. The results presented in the following are thus only directly relevant for a system with intrinsic SOC.

\subsection{Results}

\begin{figure}[t]
\centering
	\includegraphics[width=0.5\textwidth]{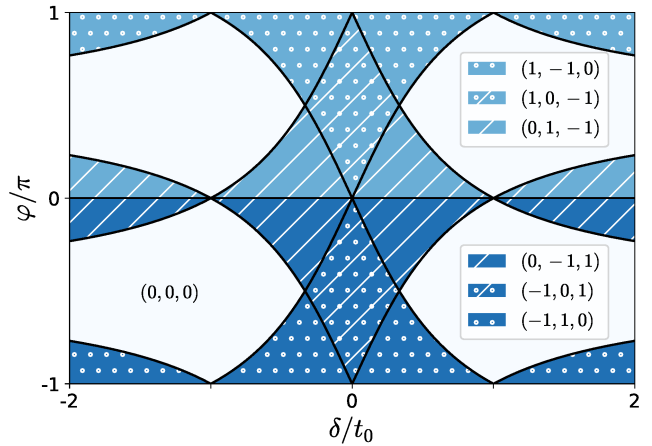}
	\caption{Diagram of the different phases characterized by the Chern numbers of the three bands. The Chern numbers are indicated as $(c_1, c_2, c_3)$, where $c_1$, $c_2$, and $c_3$ are the Chern numbers of the lowest, middle, and top bands, respectively. {The dotted  and hashed regions are topological at $1/3$ and $2/3$ filling, respectively. Darker and lighter regions of a given texture have opposite Chern numbers.}}
\label{fig:PD_chern}
\end{figure}

Because of the $C_3$ symmetry, the gap-closing-gap-opening transitions only happen at $\vb k = \vb K_{+}$ or $\vb k = \vb K_{-}$, where a pair of bands forms Dirac cones. Additionally,  when $\varphi = 0$ or $\pi$, the perfectly flat band touches one of the dispersive bands at $\vb k = \Gamma$. In Fig.~\ref{fig:PD_chern}, the different regions are delimited by lines in parameter space along which a pair of bands are touching (calculated analytically). The three Chern numbers, one for each band, are indicated in the different regions. The Chern numbers in each region were calculated numerically using the algorithm described in Ref.~\cite{fukui2005chern}.

 
\subsection{Discussion}

In Ref.~\cite{ohgushi2000spin}, the same effective model with $\delta=0$ has already been studied. They have shown that unless $\varphi=0,\pi$, the top and bottom bands have a finite Chern numbers. The same result is deduced from the $\delta=0$ line in Fig.~\ref{fig:PD_chern}.

The effect of trimerization is very different for $1/3$ and $2/3$ filling because of the  {weakly dispersing} band, in a manner reminiscent of what we observed in Sec.~\ref{sec:Z2} for the $\mathbb{Z}_2$ index.
At $1/3$ filling, the system is topologically nontrivial {in the dotted regions in Fig.~\ref{fig:PD_chern}}. For small SOC, a sufficient large {trimerization} $\delta$ (of the order of the SOC) will drive the system in a trivial phase, after a gap-closing-gap-opening transition at either $\vb K_+$ or $\vb K_-$, as is usually the case for inversion-symmetry breaking terms in Chern insulators (e.g., for the Haldane model).

However, we see that the weakly dispersing band stays topological in a large region around $\delta=0$ and $\varphi =0$. Just like in Sec.~\ref{sec:Z2}, its topological properties are robust to trimerization. {This can be seen in Fig.~\ref{fig:PD_chern}; the system is topological at $2/3$ filling in the hashed regions.}
As soon as there is an infinitesimally small SOC, lifting the degeneracy at the quadratic band touching point $\vb k = \Gamma$, the weakly dispersing band hosts a nontrivial Chern number for any values of $\delta$, except close to the fully-trimerized limit.

Finally, for Rashba SOC, we expect a similar robustness of the topology of the flat bands (as in Sec.~\ref{sec:Z2}). However, we also expect a more complex phase diagram because of the lack of $C_3$ symmetry, and the additional potential gap-closing-gap-opening transitions. We do not explicitly calculate this phase diagram here, which would depend on extra parameters (the direction of $\vu e$), in addition to $\varphi$ and $\delta$.

\section{Conclusion}
\label{sec:conclusion}
In this paper, we investigated the topological properties of electrons on the (breathing) kagome lattice with symmetry-allowed SOC between nearest neighbors. 
We drew several phase diagrams with respect to three parameters: trimerization, Rashba SOC, and intrinsic SOC.

We considered the topological phases associated with both the $\mathbb{Z}_2$ index, in the original time-reversal invariant six-band model, and the Chern numbers, in an effective spin-polarized three-band model. Interestingly, the overall topological properties (i.e., the effects of the different parameters) are similar for those two models.

We showed that the effects of Rashba SOC and the inversion symmetry-breaking term (the trimerization) is different than what is usually expected (for instance in graphene).

We showed that the pair of flat bands (or the single flat band for the spin-polarized system) is intrinsically topological. As soon as the quadratic band touching at $\Gamma$ is lifted, e.g., by an infinitesimally small SOC contribution, the now well-defined topological index of the top bands is nontrivial until a band-closing-band-opening happens close to the fully-trimerized limit $\delta = t_0$.
We stress that the {weakly dispersing} bands are topologically very robust against the perturbation (here the trimerization), unlike the other dispersive bands.

In addition, the inclusion of Rashba SOC was shown to drive the system into a topological phase by itself, in contrast to graphene where it hinders the nontrivial phase.

Finally, because of the possibility to host a nontrivial topological phase close to the fully-trimerized system, we showed how the same topological phase could be obtained from a two-band system on a {\it triangular lattice}, which could be relevant for experimental implementations.

\section{Acknowledgments}
A. B. acknowledges the Leading Graduate Course for Frontiers of Mathematical Sciences and Physics (FMSP) for the encouragement of the present paper. 

N. N. was supported by JSPS KAKENHI Grant Number JP26103006, JP18H03676, and ImPACT Program of Council for Science, Technology and Innovation (Cabinet office, Government of Japan), and JST CREST Grant Numbers JPMJCR16F1, and JST CREST Grant Number JPMJCR1874, Japan.

\bibliographystyle{apsrev4-1}

%

\end{document}